\documentclass[aps,prb,twocolumn,showpacs,floats]{revtex4}
\usepackage{latexsym}
\usepackage{amsmath}
\usepackage{amssymb}
\usepackage{bm}
\usepackage{wasysym}
\usepackage[dvips]{color}
\usepackage{graphicx}
\usepackage{graphicx}
\usepackage{subfigure}
\usepackage{epsfig}

\def\up{\uparrow}
\def\down{\downarrow}

\newcommand{\veps}{\varepsilon}
\newcommand{\bfk}{{\bf k}}

\def\mbfk{\mathbf{k}}

\begin{document}

\title{Anderson Impurity in the Bulk of 3D Topological Insulators:\\
       II. The Strong Coupling Regime}

\author{Igor Kuzmenko$^1$, Tetyana Kuzmenko$^{1}$,
         Yshai Avishai$^{1,2}$ and Tai Kai Ng$^3$}
 \affiliation{\footnotesize
   $^1$Department of Physics, Ben-Gurion University of
   the Negev Beer-Sheva, Israel \\
 $^2$NYU-Shanghai, Pudong, Shanghai, China\\
   $^3$Department of Physics, Hong Kong University of Science and
   Technology, Kowloon, Hong Kong}

\begin{abstract}
 Electron scattering off an Anderson impurity immersed in the bulk of
a 3D topological insulator is studied in
the strong coupling regime, where the temperature $T$ is
lower than the Kondo temperature $T_K$. 
The system displays
either a self-screened Kondo effect, or a Kondo
effect with SO(3) or SO(4) dynamical symmetries. 
Low temperature Kondo scattering
for systems with SO(3) symmetry displays the behavior of
a singular Fermi liquid, an elusive property that so far has been 
observed only in tunneling experiments.  
This is demonstrated through the singular behavior as $T \to 0$ of the 
specific heat, magnetic susceptibility
and impurity resistivity, 
that are calculated  using well known 
(slightly adapted) conformal field theory techniques. 
Quite generally, the low temperature
dependence of some of these observables displays
a remarkable distinction between the SO(n=3,4)
Kondo effect, compared with the standard SU(2) one. 
\end{abstract}


\maketitle

\section {Introduction}
  \label{sec-intro}

In this work we continue our study of a system composed
of an Anderson impurity $d$ immersed in a 3D topological
insulator (3DTI)
with an ``inverted-Mexican-hat" band dispersion around
the $\Gamma$-point \cite{23,TI3D-paper1,24,top-ins-3D-10,%
3D-TI-QD-12,InvMexHat-PRB-2013}.
For the sake of self-consistency let us very briefly recapitulate the peculiar feature 
of the pertinent system: 
Due to the special band structure,  the ensuing Kondo effect
(KE) is profoundly distinct from its
metallic analog in normal metals,
because in addition to the original Anderson impurity $d$, there is 
an in-gap bound state (henceforth denoted as an $f$ impurity),
that is formed as a result of potential scattering\cite{23,TI3D-paper1}. 
The $d$ and $f$ impurities form a ``composite quantum impurity" (CQI) that 
turns the pertinent Kondo physics much richer. 
The reason for that is as follows: When isolated, the CQI
can host two electrons that are found in singlet or triplet states
with corresponding energies $E_S$ and $E_T$.
Due to hybridization of the localized electrons in the CQI with the band electrons,
the levels $E_S$ and $E_T$ are renormalized with decreasing bandwidth
{\it albeit with different rates} \cite{TI3D-paper1}. As a result, there is either a
self-screened KE, or a KE with SO(3) or SO(4)
dynamical symmetry (depending on whether at the end of renormalization,
$E_T$ lies
above, below or coincides with $E_S$
\cite{23,TI3D-paper1}.

In our previous work\cite{TI3D-paper1}, we have analyzed
the pertinent Kondo physics {\it in the weak coupling regime}
using perturbative RG analysis techniques.
The goal of the present work is to perform an analysis of the
Kondo scattering with SO(3) and SO(4) dynamical symmetry
in the {\it strong coupling regime} $T<T_K$. The main physical
motivation is to elucidate the occurrence of singular Fermi
liquid in the SO(3) symmetric sector, that is exposed
{\it only in the strong coupling regime}. More
than a decade ago it has been shown that the traditional
classification of quantum impurity models into Fermi liquids
and non Fermi liquids should be modified in such a way that
one has to distinguish between regular Fermi liquids
and singular Fermi liquids (SFL)\cite{Coleman-bethe-anzats-prb03,%
Coleman-Paul, Coleman-bethe-anzats-prb05}. 
The former case is exemplified by the standard
SU(2) Kondo model where electrons are scattered
from a magnetic impurity of spin $S=\frac{1}{2}$ 
and at zero temperature the impurity is {\it fully screened}. 
This makes it possible to describe the system in terms
of Nozi\`eres Fermi-liquid picture. On the other hand,
it was shown in Ref.~\cite{Coleman-bethe-anzats-prb05}
that when the impurity is {\it under-screened},
the corresponding Fermi liquid is singular. Practically,
it implies that the density of states is (logarithmically)
singular at the Fermi energy.

As is already stressed, this manifestation of SFL occurs only
at low temperature $T<T_K$, and that requires 
the calculations to be carried out in the strong coupling
regime,  where perturbation theory is not applicable.
One need to resort to other approaches, such as Bethe
ansatz or conformal field theory (CFT, that is employed
here). Calculated experimental observables for SO(3)
and SO(4) symmetric Kondo effect include the impurity
contribution to the temperature dependence of
the specific heat, magnetic susceptibility and resistivity. 
Comparing these results with  those of the standard
SU(2) Kondo effect, we indeed elucidate the remarkable
SFL nature of the  KE
for the SO(3) symmetric sector.

Experimentally, under-screened Kondo effect (USKE)
has been observed in tunneling transport measurements
through molecular transistors \cite{USKE-exp-PRL2009,%
USKE-exp-Sci2010,USKE-exp-PRL2015}. It is shown
therein that the tunneling conductance $G$ approaches
the unitary limit when the temperature $T$ tends to zero,
but the derivative $dG/dT$ is divergent. The possibility
of detecting USKE in Kondo scattering in the bulk of
metals is hampered by the fact that quantum impurities
with high spin give rise to Coqblin-Schrieffer type of
Kondo scattering which eventually flows to a regular
Fermi liquid fixed point \cite{Hewson-book}.
In this work we overcome this obstacle and demonstrate
the feasibility of observing USKE in bulk materials
employing the occurrence of composite impurities.

The organization of the paper is facilitated by
the fact that the model and the relevant starting
Kondo Hamiltonian as well as the Kondo temperature
have already been derived  in our previous work 
(starting from the single impurity Anderson Hamiltonian).  
Therefore, in Sec. \ref{sec-HK} we start right away by
writing down the Kondo Hamiltonian $H_K$
describing low-energy exchange scattering of
the band electrons by the $d$ and $f$ impurities
as well as an  exchange interactions between
the $d$ and $f$ impurities. The Hamiltonian $H_K$
possesses different dynamical symmetries, SO(3)
or SO(4), within various energy domains. The local
density of states (DOS) of the TI with  magnetic
impurity immersed in it is calculated in Sec.
\ref{sec-KE-spin-S}. Kondo Interaction of
the "dressed" Fermi liquid (formed by the band
electrons and the $d$ impurity) with the $f$-impurity
is discussed in Sec.~\ref{sec-second-TK}.
In Sec. \ref{sec-specific-heat} we study and
calculate the temperature dependence of
the impurity contribution to the specific heat,
while the magnetic susceptibility of the impurity
is considered in Sec. \ref{sec-susceptibility}. Sec. \ref{sec-resist}
is devoted to the calculations of electric resistivity. The results
are then summarized in Sec. \ref{sec-conclusions}.
Some technical details are relegated to the Appendices.
In Appendix \ref{append-model} we describe the ground
state of the isolated composite impurity.
Weak coupling renormalization of the exchange interaction
strength of the $f$-impurity with the band is considered
in Appendix \ref{append-scaling-SO(4)}.

\section{Kondo Hamiltonian}
  \label{sec-HK}

As it is shown in our previous work \cite{TI3D-paper1},
the lowest-energy states of the isolated
CQI  are the singlet
and triplet states with one electron on the $d$-level and
a second electron on the $f$-level. Electron tunneling between
the $d$-level and the band modifies the number of electrons
in the CQI. Integrating out high energy states from the band
edges renormalizes the singlet and triplet levels until charge
fluctuations are quenched and one arrives at the  local
moment regime. In this regime, the Schrieffer-Wolff
transformation is used to map the Anderson Hamiltonian
onto an effective Hamiltonian $H=H_0+H_K$.
Here the first term, $H_0$, describes electrons in the bulk of the TI,
\begin{subequations}
\begin{eqnarray}
  H_0 &=&
  \sum_{\nu\sigma{\bf{k}}}
  \nu\veps_{k}
  \gamma_{\nu{\bf{k}}\sigma}^{\dag}
  \gamma_{\nu{\bf{k}}\sigma},
  \label{H-ti-def}
\end{eqnarray}
where $\nu=\pm 1$ denotes the conduction and valence
band.
\begin{equation}
  \label{disp}
  \veps_{k}=
  \sqrt{M_{k}^2+(\hbar v k)^2},
  \ \ \
  M_{k}=m v^2-B \hbar^2k^2,
\end{equation}
is the band dispersion.
  \label{subeqs-H-lead-def}
\end{subequations}
Accordingly, $\veps_k$ is gapped and the insulator is
topological for $Bm>0$. For $Bm>1/2$ (assumed hereafter),
the band dispersion has an ``inverted-Mexican-hat'' form
with dispersion minimum at a surface of nonzero wave-vector
${\mathbf{q}}$'s, with
\begin{eqnarray}
  \veps_q ~=~
  \frac{v^2}{B}~
  \sqrt{Bm-\frac{1}{4}},
  \ \ \
  q ~=~
  \frac{v}{\hbar v}~
  \sqrt{Bm-\frac{1}{2}},
\end{eqnarray}
where $q=|{\mathbf{q}}|$.

The Kondo Hamiltonian $H_K$ assumes the following form
\cite{TI3D-paper1}:
\begin{equation}
  \label{sw2k}
  H_K =
  J_d
  \big(
      \mathbf{S}_d
      \cdot
      {\mathbf{s}}
  \big)+
  J_f
  \big(
      \mathbf{S}_f
      \cdot
      {\mathbf{s}}
  \big)+
  J_{df}
  \big(
      \mathbf{S}_d
      \cdot
      \mathbf{S}_f
  \big),
\end{equation}
where ${\mathbf{S}}_{d}$ or ${\mathbf{S}}_{f}$ is the localized
spin of the $d$- or $f$-impurity, ${\mathbf{s}}$ is the spin
operator of the band electrons,
$$
  {\mathbf{s}}=
  \frac{1}{2}~
  \sum_{\nu\mathbf{k}\sigma,\nu'\mathbf{k}'\sigma'}
  \left(
       \gamma^{\dag}_{\nu\mathbf{k}\sigma}
       {\boldsymbol\tau}_{\sigma\sigma'}
       \gamma_{\nu'\mathbf{k}'\sigma'}
  \right),
$$
$\hat{\boldsymbol\tau}=(\hat\tau^x,\hat\tau^y,\hat\tau^z)$ is the
vector of the Pauli matrices. The couplings $J_K$, $J_f$ and
$J_{df}$ are explicitly given as \cite{TI3D-paper1},
\begin{subequations}
\begin{eqnarray}
  J_d &\sim&
  \frac{2V_d^2}{\epsilon_F-\epsilon_d},
  \ \ \ \ \
  J_f ~\sim~
  \frac{J_d \beta_f^2}{4},
  \label{JK-def}
  \\
  J_{df} &\sim&
  \beta_f^2\Delta_f
  \left(
       1-
       \frac{2V_d^2\rho_c}{\Delta_f}~
       \frac{\sqrt{D_0}-\sqrt{|\epsilon_f|}}
            {\sqrt{\varepsilon_0}}
  \right),
  \label{tilde-J-def}
\end{eqnarray}
  \label{subeqs-J-def}
\end{subequations}
where
$$
  \beta_{f} ~=~
  \frac{\sqrt{2}~V_{df}}{\Delta_f},
  \ \ \ \ \
  \Delta_f ~=~
  \epsilon_f-
  \epsilon_d+
  U_f.
$$
Here $\epsilon_d$ and $\epsilon_f$ are single electron
energies of the $d$- and $f$-impurities, $V_d$ is
hybridization rate of the $d$-impurity and the band and
$V_{df}$ is the hybridization of the $d$- and $f$-impurities
[see Fig. \ref{Fig-lead-d-f} below],
and $U_f$ is the Coulomb blockade parameter for
the $f$-impurity. The Coulomb blockade $U_d$ of
the $d$-impurity is assume to be infinity \cite{TI3D-paper1}.
Notice that, generally, ${J_f}\ll{J_d}$ \cite{TI3D-paper1,24}.

\noindent
The low-energy physics of the model depends on $J_{df}$. More
concretely, there are three different regimes for $J_{df}$
determining the different ground states (GS)
\cite{TI3D-paper1,Affleck-Ludwig-Jones-prb95}:
\begin{itemize}
\item When $J_{df}=0$, the impurities are decoupled, and
       Kondo scattering of the band electrons is determined
      mostly by the hybridization of the $d$-impurity
      with the band, whereas the $f$-impurity can be considered
      as an isolated magnetic moment (i.e., it is not coupled to
      the band or to the $d$-impurity). In this  case, there is
      standard Kondo effect with full screening of the spin of
      the $d$-impurity. The exchange interaction of the $f$-impurity
      with the band manifests itself just at temperature below
      Kondo temperature. Therefore,  $J_f$ enters into consideration only 
      for temperatures below the Kondo temperature.

\item When $J_{df}<0$ with $|J_{df}|\gg{J}_{d}$,
      the ground state of the CQI is a triplet and the singlet state
     is highly excited. In this case, band electrons are scattered from a 
     CQI with spin $S=1$, and there is an  USKE of the magnetic impurity.
      Since the magnetic moments of the $d$-
      and $f$-impurities in the triplet ground state are parallel
       to one another, the exchange interaction strength $J_f$ of
      the $f$-impurity with the lead gives rise to slight modification
      of $J_d$ \cite{TI3D-paper1}, (that is the exchange interaction strength
      of the $d$ impurity with the band electrons). Therefore we assume
      $J_f=0$ in this regime.

\item When $J_{df}>0$ with $J_{df}\gg{J}_{d}$,
      the ground state of the CQI is singlet and the triplet state
      has high excitation energy. In this case, there is no Kondo
      effect.
\end{itemize}

Before presenting our calculations pertaining to physical observables in our system, let us
recapitulate the nature of the GS, specific heat and magnetic
susceptibility for the ordinary case of a magnetic impurity immersed in an {\it ordinary
metal} [\onlinecite{Coleman-prb07}].   For
the fully screened (FS) and the under screened (US) cases of
the KE they are listed in the table below. Recall that the GS of the FSKE is
a Fermi a liquid (FL) while that of the USKE turns out to
be a SFL. Explicitly,
\begin{center}
\begin{tabular}{|c||c|c|c|c|}
\hline KE & GS & $C_V$ & $\chi$
\\
\hline \hline FS & FL & $\sim{T}$ & $\sim{T}_{K}^{-1}$
\\
\hline US & SFL & $\sim{T}\ln^{4}\big(\frac{T}{T_K}\big)$ &
$\sim{T}^{-1}\ln^{-2}\big(\frac{T}{T_K}\big)$
\\
\hline
\end{tabular}
\end{center}
In the following, we will consider physical properties of the
system in the strong coupling regime (when the temperature is
below the Kondo temperature) for the cases $J_{df}<0$ and
$J_{df}=0$, in turn.

\section{Density of States}
  \label{sec-KE-spin-S}

\begin{figure}[htb]
\centering
\includegraphics[width=60 mm,angle=0]{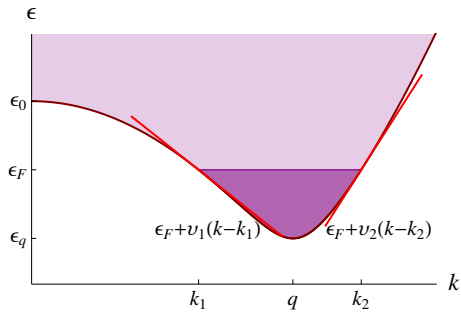}
\caption{\footnotesize
   {\color{blue}(Color online)}
  Dispersion $\veps_k$, Eq. (\ref{disp}). $k_i$ ($i=1,2$) are
  two solutions of the equation $\veps_k=\epsilon_F$.
  $v_i=v_{k_i}$ are the Fermi velocities at the inner or
  outer Fermi surfaces ($k=k_1$ or $k_2$).}
 \label{Fig-disp}
\end{figure}

In this section we apply the already known 
Bethe ansatz expression for the scattering phase shift and 
write down the density of states (DOS) that is peculiar 
to the system of spin S CQI immersed in a material with inverse Mexican hat 
band structure. Specifically, we focus on the low-energy 
spectrum of the system where it is
justified to approximate the dispersion (\ref{disp}) by linear
expressions in $k-k_i$ ($i=1,2$), where $k_1$ and $k_2$ are two
solutions of the equation $\veps_k=\epsilon_F$ (see Fig.~\ref{Fig-disp}).
The DOS $N_S(\omega)$ of the Fermi
gas near the impurity position is expressed in terms of
the phase shift $\delta_S(\omega)$ through the Friedel
sum rule \cite{Coleman-bethe-anzats-prb05},
\begin{eqnarray}
  N_S(\omega) &=&
  \frac{1}{\pi}~
  \frac{d\delta_S(\omega)}{d\omega},
  \label{DOS-vs-delta}
\end{eqnarray}
where $S=\frac{1}{2}$ or $1$ is the impurity spin.
Calculations based on the Bethe anzats
\cite{Coleman-bethe-anzats-prb05} applied for the full
screened $S=\tfrac{1}{2}$ and the under-screened $S=1$
Kondo effect, yield the following expression for
the phase shift,
\begin{eqnarray}
  \delta_{S}(\omega) &=&
  \frac{\pi}{2}+
  \frac{1}{2i}~
  \ln
  \left(
       \frac{\Gamma\big(S+\frac{1}{2}+\frac{i}{\pi}\ln\big(\frac{\omega}{T_K}\big)\big)}
            {\Gamma\big(S+\frac{1}{2}-\frac{i}{\pi}\ln\big(\frac{\omega}{T_K}\big)\big)}~
  \right)+
  \nonumber \\ && +
  \frac{1}{2i}~
  \ln
  \left(
       \frac{\Gamma\big(S-\frac{i}{\pi}\ln\big(\frac{\omega}{T_K}\big)\big)}
            {\Gamma\big(S+\frac{i}{\pi}\ln\big(\frac{\omega}{T_K}\big)\big)}
  \right).
  \label{delta-BA}
\end{eqnarray}
For $S=\frac{1}{2}$, the above expression takes the form,
\begin{eqnarray}
  \delta_{\frac{1}{2}}(\omega) &=&
  \frac{\pi}{2}-
  \arctan\Big(\frac{\omega}{T_K}\Big),
  \label{delta-S=1/2}
\end{eqnarray}
whereas for $S=1$ and $\omega\ll{T}_{K}$, $\delta_S(\omega)$
has the asymptotic expression,
\begin{eqnarray}
  \delta_{1}(\omega) &=&
  \frac{\pi}{2}
  \bigg\{
       1+
       \frac{1}
            {2
             \ln
             \big(
                 \frac{T_K}{\omega}
             \big)}
  \bigg\}.
  \label{delta-S=1-asymptot}
\end{eqnarray}
Note that $\delta_{1}(\omega)$ demonstrates a singular
behavior near the point $\omega=0$, whereas
$\delta_{\frac{1}{2}}(\omega)$ is regular.
Differentiating $\delta_S$ (\ref{delta-BA}) gives the following
expression for the DOS \cite{Coleman-bethe-anzats-prb05}:
\begin{eqnarray}
  N_S(\omega) &=&
  \frac{1}{2\pi\omega}~
  {\mathrm{Re}}
  \Bigg[
       \beta
       \bigg(
           S+
           \frac{i}{\pi}~
           \ln\frac{\omega}{T_K}
       \bigg)
  \Bigg].
  \label{DOS-Bethe-Anzats}
\end{eqnarray}
Here $S=1/2$ or $1$ is the spin of the impurity, and the function
$\beta(x)$ is defined as
\begin{eqnarray}
  \beta(x) &=&
  \frac{1}{2}~
  \Bigg\{
       \psi
       \bigg(
            \frac{x+1}{2}
       \bigg)-
       \psi
       \bigg(
            \frac{x}{2}
       \bigg)
  \Bigg\},
\end{eqnarray}
where $\psi(x)$ is the digamma function,
\begin{eqnarray*}
  \psi(x) &=&
  \frac{d\ln\Gamma(x)}{dx} ~=~
  \frac{\Gamma'(x)}{\Gamma(x)}.
\end{eqnarray*}
\begin{figure}[htb]
\centering
\includegraphics[width=60 mm,angle=0]{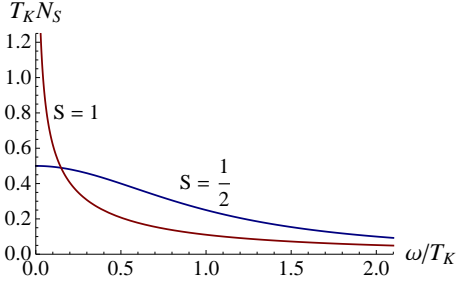}
\caption{\footnotesize
   {\color{blue}(Color online)}
  DOS Eq.~(\ref{DOS-Bethe-Anzats}) for the SO(4) KE
  (blue curve) and the SO(3) KE (red curve).}
 \label{Fig-DOS}
\end{figure}
The DOS Eq.~(\ref{DOS-Bethe-Anzats}) is shown in Fig. \ref{Fig-DOS}
for $S=\frac{1}{2}$ and $S=1$. It is seen that the DOS for $S=1$
is singular. As a result the conventional Fermi liquid expansion
of the phase shift can not be carried out
\cite{Coleman-bethe-anzats-prb05}.
The origin of this singularity is the non-analytic behavior
of the phase shift $\delta_{1}(\omega)$, eq. (\ref{delta-BA}),
near the point $\omega=0$.

\section{Kondo Interaction of the Fermi Liquid with the $f$-Impurity}
  \label{sec-second-TK}
When $|J_{df}|\ll{T}_{K_4}$ [where $T_{K_4}$ is the Kondo
temperature for the SO(4) KE, see Ref.~ \cite{TI3D-paper1}
and Eq. (\ref{TK4-def}) in Appendix \ref{append-scaling-SO(4)}],
the $d$-impurity and the conduction band electrons form
a singlet state, and the system composed of the $d$-impurity
{\it and the band electrons} forms a local Fermi liquid that can
be described within Nozi\`eres Fermi liquid theory\cite{Noz}.
The $f$-impurity is coupled to this local Fermi liquid through
an effective exchange Hamiltonian,
\begin{eqnarray}
  H_{K}^{(2)} &=&
  \tilde{J}_{f}
  \big(
      {\mathbf{S}}_{f}
      \cdot
      {\mathbf{s}}
  \big),
  \label{HK-2}
\end{eqnarray}
where
\begin{eqnarray}
  \tilde{J}_{f} &=&
  \frac{J_{df}(T_{K_4})}{\rho_0}.
  \label{tilde-Jf}
\end{eqnarray}
The density of states of the local Fermi liquid is,
\begin{eqnarray}
  \tilde\rho(\epsilon) &=&
  \frac{T_{K_4}}
       {\big(\epsilon-\epsilon_F\big)^{2}+
        T_{K_4}^{2}}.
  \label{DOS-tilde}
\end{eqnarray}

The dimensionless coupling $\tilde{j}_{f}$ is,
\begin{eqnarray}
  \tilde{j}_{f} &=&
  \tilde{J}_{f}
  \tilde\rho(\epsilon_F)
  ~\sim~
  \frac{j_f \veps_q}{T_{K_4}}
  ~\sim
  \nonumber \\ &\sim&
  j_f~
  \frac{\veps_q}{D_{ii}}~
  \exp
  \bigg(
       \frac{1}{j_d}
  \bigg),
  \label{tilde-jf-def}
\end{eqnarray}
where we take into account that $\rho_{\mathrm{c}}\sim1/\veps_q$.
The scaling equation for $\tilde{j}_{f}$ is,
\begin{eqnarray}
  \frac{\partial \tilde{j}_{f}}
       {\partial \ln D}
  &=&
  -\tilde{j}_{f}^{2}.
  \label{eq-tilde-jf}
\end{eqnarray}
The initial value $\tilde{j}_{f}(T_{K_4})$ of $\tilde{j}_{f}(D)$
is given by eq. (\ref{tilde-jf-def}).
The solution of the scaling equation (\ref{eq-tilde-jf}) is,
\begin{eqnarray}
  \tilde{j}_{f}(D) &=&
  \frac{1}
       {\displaystyle
        \ln
        \bigg(
             \frac{D}{T_{K}^{(2)}}
        \bigg)},
  \label{tilde-jf-solution}
\end{eqnarray}
where the scaling invariant, the second Kondo temperature, is
\begin{eqnarray}
  T_{K}^{(2)} =
  T_{K_4}
  \exp
  \Bigg\{
       -\frac{1}{j_f}~
       \frac{D_{ii}}{\veps_q}~
       \exp
       \bigg(
            -\frac{1}{j_d}
       \bigg)
  \Bigg\}.
  \label{TK-second}
\end{eqnarray}

For $\epsilon_d=-80\epsilon_q$, $\veps_0=24\veps_q$,
$\epsilon_F=2\veps_q$, $j=0.08$ and $j_f=0.025$, we get
$T_{K_4}=0.15\veps_q$ and $T_{K}^{(2)}=0.0021T_{K_4}$.

Having set up the calculation framework we turn now to elucidate
numerous physical observable. These include specific heat,
magnetic susceptibility and impurity resistivity. Since
$T_K^{(2)}$ is very small, we shall restrict ourselves to
the temperature regime $T\gg{T}_{K}^{(2)}$ in the following. For
the SO(4) KE, the temperature is in the range
${T}_{K_4}>{T}\gg{T}_{K}^{(2)}$. For the SO(3) KE, we consider
$T<T_{K_3}$, where $T_{K_3}$ and
$T_{K_4}$ are the Kondo temperatures for the SO(3) and SO(4) KE,
respectively \cite{TI3D-paper1}. For the SO(3) KE, we employ CFT
techniques, whereas for the SO(4) KE, we will apply CFT techniques
for the interaction between the $d$-impurity and the conduction band
electrons, and then we employ poor man's scaling formalism to take
into account the interaction between the local Fermi liquid and
the $f$-impurity.

\section{Specific Heat}
  \label{sec-specific-heat}

The first observable to be calculated is the specific heat of
the TI within which the  CQI is immersed. Its definition reads,
\begin{eqnarray}
  C_{V}^{(0)} &=&
  \frac{\partial}{\partial T}
  \int{d\epsilon}~
  \epsilon
  f(\epsilon)
  N_S\big(|\epsilon-\epsilon_F|\big),
  \label{specific-heat-def}
\end{eqnarray}
where $f(\epsilon)$ is the Fermi function.
Noting  that
$$
  \frac{\partial f(\epsilon)}
       {\partial T}
  ~=~
  \frac{\epsilon-\epsilon_F}
       {4T^2
        \cosh^2
        \big(
            \frac{\epsilon-\epsilon_F}{2T}
        \big)},
$$
we have
\begin{eqnarray}
  C_{V}^{(0)} &=&
  4 T
  \int\limits_{0}^{\infty}
  \frac{N_S(2Tx)~x^2~dx}
       {\cosh^2(x)}.
  \label{specific-heat}
\end{eqnarray}

For the SO(3) KE, the specific heat is ${C}_{V}={C}_{V}^{(0)}$
[given by Eq.~(\ref{specific-heat})]. For the SO(4) KE, there is
contribution to the specific heat due to the $f$-impurity.
This contribution can be written as,
\begin{eqnarray}
  \delta{C}_{V} &=&
  \frac{3\pi^2}{4}~
  \tilde{j}_{f}^{4}(T),
  \label{specific-heat-correction}
\end{eqnarray}
where $\tilde{j}_{f}(T)$ is given by Eq. (\ref{tilde-jf-solution}).
The total specific heat for the SO(4) KE is
${C}_{V}={C}_{V}^{(0)}+\delta{C}_{V}$.

\begin{figure}[htb]
\centering
\includegraphics[width=60 mm,angle=0]{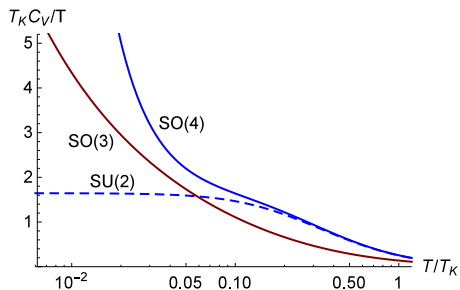}
\caption{\footnotesize
  {\color{blue}(Color online)}
  Ratio $C_V/T$ [where $C_V$ is the specific heat
  (\ref{specific-heat})] for the SO(4), SO(3) and SU(2) KE
  (solid blue, solid red and dashed blue curves). Here
  $T_K=T_{K_4}$ for the SO(4) and SU(2) KE or
  $T_{K_3}$ for the SO(3) KE.}
  \label{Fig-CV-CVT}
\end{figure}

The ratio $C_V/T$ [where $C_V$ is the specific heat
(\ref{specific-heat})] is shown in Fig. \ref{Fig-CV-CVT}
for the SO(4) and SO(3) KE.  In this case, the Kondo
temperature is $T_{K_4}$. For comparison,
the specific heat for the SU(2) KE ( i.e., for the case when
the $f$-impurity is absent), is shown as well.
 It is seen that for the SU(2) KE,
$C_V/T$ saturates to constant as $T \to 0$, in agreement
with the prediction of Fermi liquid theory. The ratio
$C_V/T$ for the SO(4) KE is close to that for the SU(2)
KE for $T$ close to $T_{K_4}$. For low temperatures,
$C_{V}/T$ for the SO(4) and SU(2) KE deviate
substantially from one another. This is result is due
to the exchange interaction of the $f$-impurity with
the local Fermi liquid. For the SO(3) KE, the ratio
$C_V/T$ diverges indicating the feature of
a SFL behavior \cite{Coleman-bethe-anzats-prb05}.

\section{Magnetic Susceptibility}
  \label{sec-susceptibility}

When $J_{df}<0$ and the ground state of the isolated CQI is $S=1$,
the magnetic susceptibility is,
\begin{eqnarray}
  \chi(T) =
  2\mu_B^2
  \int{d\epsilon}~
  N_1\big(|\epsilon-\epsilon_F|\big)~
  \bigg(
       -\frac{\partial f(\epsilon)}{\partial \epsilon}
  \bigg),
  \label{susceptibility-spin1-def}
\end{eqnarray}
where $N_1(\epsilon)$ is given by Eq. (\ref{DOS-Bethe-Anzats})
with $S=1$. Taking into account that
\begin{eqnarray}
  -\frac{\partial f(\epsilon)}{\partial \epsilon}
  &=&
  \frac{1}
       {4T
        \cosh^2
        \big(
            \frac{\epsilon-\epsilon_F}{2T}
        \big)},
  \label{df-de-def}
\end{eqnarray}
we can write,
\begin{eqnarray}
  \chi(T) &=&
  \frac{2 \chi_0 T_K}{T}
  \int
  \frac{N_1(|\epsilon-\epsilon_F|)~
        d\epsilon}
       {\cosh^2
        \big(
            \frac{\epsilon-\epsilon_F}{2T}
        \big)},
  \label{susceptibility-spin1-def1}
\end{eqnarray}
where
\begin{eqnarray}
  \chi_0 &=&
  \frac{\mu_B^2}{T_K}.
  \label{chi0-def}
\end{eqnarray}
For the case $J_{df}=0$ (see Eq.~(\ref{subeqs-J-def})),
the susceptibility is calculated in the following way:
First, we employ the CFT technique to calculate
the susceptibility for $\tilde{j}_{f}=0$ (see
Eq.~(\ref{tilde-jf-solution})). Within this approximation,
the CQI splits into two noninteracting
impurities, one of them is coupled to the band electrons,
and the other is fully decoupled. In this case,
the susceptibility is given by,
\begin{eqnarray}
  \chi^{(0)}(T) &=&
  2\mu_B^2
  \int{d\epsilon}~
  N_{\frac{1}{2}}\big(|\epsilon-\epsilon_F|\big)~
  \bigg(
       -\frac{\partial f(\epsilon)}{\partial \epsilon}
  \bigg)+
  \nonumber \\ && +
  \frac{\mu_B^2}{T},
  \label{susceptibility-spin1/2-def}
\end{eqnarray}
where $N_{\frac{1}{2}}(\epsilon)$ is given by Eq.
(\ref{DOS-Bethe-Anzats}) with $S=\frac{1}{2}$. The second term,
$\mu_B^2/T$, is the susceptibility of the $f$-impurity. Taking  into
account Eq. (\ref{df-de-def}), we can write,
\begin{eqnarray}
  \chi^{(0)}(T) =
  \frac{2 \chi_0 T_K}{T}
  \int
  \frac{N_{\frac{1}{2}}\big(|\epsilon-\epsilon_F|)~
        d\epsilon}
       {\cosh^2
        \big(
            \frac{\epsilon-\epsilon_F}{2T}
        \big)}+
  \frac{\mu_B^2}{T},
  \label{susceptibility-spin1/2-def1}
\end{eqnarray}
where $\chi_0$ is given by Eq. (\ref{chi0-def}).
The contribution to the susceptibility due to the interaction
of the local Fermi liquid with the $f$-impurity is,
\begin{eqnarray}
  \delta\chi(T) &=&
  -\frac{\chi_0 T_K}{T}~
  \tilde{j}_{f}(T),
  \label{susceptibility-spin1/2-correction}
\end{eqnarray}
where $\tilde{j}_{f}(T)$ is given by Eq. (\ref{tilde-jf-solution}).
The total susceptibility for the SO(4) KE is
\begin{eqnarray}
  \chi(T) &=&
  \chi^{(0)}(T)+
  \delta\chi(T).
  \label{susceptibility-SO(4)}
\end{eqnarray}
\begin{figure}[htb]
\centering
\includegraphics[width=60 mm,angle=0]{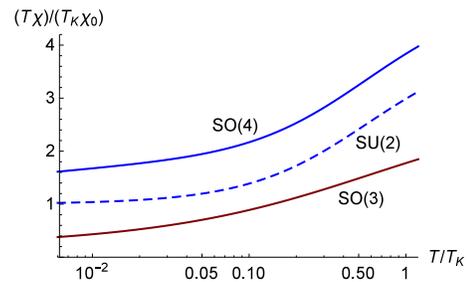}
  \label{Fig-suscept-b}
\caption{\footnotesize
  {\color{blue}(Color online)}
  The quantity $T\chi(T)$ for the SO(3), SU(2) and the SO(4) KE
  is displayed as a function of temperature.
  The saturation as $T \to 0$ for the SU(2) and SO(4) KE is
  reminiscent of the Curie law. On the other hand, for the SO(3) KE,
  $T\chi(T) \to 0$.  }
 \label{Fig-suscept}
\end{figure}
The functions $T\chi(T)$ for the
SO(3), SU(2) and SO(4) KE are shown in Fig. \ref{Fig-suscept}.
The SU(2) KE corresponds to
the case when the $f$-impurity is absent. For this case, the Kondo
temperature is $T_{K_4}$.
It is seen that for the SU(2) or SO(4) KE, $T\chi(T)\to{T_K}\chi_0$ for $T\to0$.
For the SO(3) KE, $T\chi(T)$ vanishes as $\ln^{-2}(T/T_K)$ when
$T\to0$, which is a manifestation of a singular Fermi liquid
fixed point.

\section{Impurity Resistivity}
  \label{sec-resist}
  \noindent

In this section we will calculate the impurity contribution
to the resistivity. Here, the peculiar band structure of
the TI plays a central role. The calculation method of
the impurity contribution to the resistivity depends on
the underlying symmetry. The CFT technique used
above to derive the specific heat and magnetic field
will be employed to derive the resistivity of the KE with
the SO(3) symmetry. On the other hand, for the KE with
the SO(4) symmetry, we apply a somewhat different
approach: First, we apply the CFT technique to derive
the resistivity for the case $\tilde{j}_{f}=0$ (see
Eq. (\ref{tilde-jf-solution})). In the next step, we apply
the perturbative RG (poor man's scaling approach)
to get the correction to the resistivity due to the interaction
of the $f$-impurity with the local Fermi liquid.

\noindent
For $\tilde{j}_{f}=0$, the impurity resistivity can be
written as \cite{Hewson-book}
\begin{subequations}
\begin{eqnarray}
  \rho^{(0)}(T) &=&
  \frac{1}{\sigma^{(0)}(T)}.
  \label{resist-def}
\end{eqnarray}
Here the conductivity $\sigma^{(0)}(T)$ is
\begin{eqnarray}
  \sigma^{(0)}(T) =
  \frac{2e^2}{3}
  \int\frac{d^3\mbfk}{\big(2\pi\big)^3}
  \bigg[
       -\frac{\partial f(\epsilon_k)}{\partial\epsilon_k}
  \bigg]
  v_k^2
  \tau_{\mathrm{tr}}(k),
  \label{conduct-def}
\end{eqnarray}
  \label{subeqs-resist-conduct}
\end{subequations}
where $f(\epsilon)$ is the Fermi-Dirac distribution,
and $v_k=|{\mathbf{v}}_{\mbfk}|$ is the group velocity,
\begin{eqnarray*}
  {\mathbf{v}}_{\mbfk} &=&
  \frac{1}{\hbar}~
  \frac{\partial\epsilon_k}
       {\partial\mbfk} =
  \frac{1}{\hbar}~
  \hat{\bfk}
  \frac{\partial\veps_k}{\partial k}.
\end{eqnarray*}
For the dispersion (\ref{disp}), ${\mathbf{v}}_{\mbfk}$ is
given explicitly by,
\begin{eqnarray}
  {\mathbf{v}}_{\mbfk} &=&
  \frac{\veps_0}{\veps_k}~
  \big(2Bm-1\big)~
  \bigg(
       \frac{k^2}{q^2}-1
  \bigg)~
  \frac{\hbar\mbfk}{mv},
  \label{group-velocity}
\end{eqnarray}
where
$$
  \veps_0 ~=~
  mv^2.
$$
The resistivity is directly related to the inverse of
the transport relaxation time $\tau_{\mathrm{tr}}(k)$
that is expressible in terms of the phase shifts $\eta_l(k)$
(corresponding to angular moment $l$) appearing in
the partial wave expansion of the electron scattering
from the impurity [see Eq. (2.20) in Ref.\cite{Hewson-book}].
Explicitly,
\begin{eqnarray}
  \frac{1}{\tau_{\mathrm{tr}}(k)} =
  \frac{2c_{\mathrm{imp}}}
       {\pi\hbar\nu(k)}
  \sum_{l=1}^{\infty}
  l
  \sin^2  \big[
      \eta_l(k)-
      \eta_{l-1}(k)
  \big] ,
  \label{relax-time-phases}
\end{eqnarray}
where $\nu(k)$ is the bare density of states (DOS),
\begin{eqnarray}
  \nu(k) &=&
  \frac{1}{L^3}
  \sum_{\mbfk'}
  \delta(\epsilon_k-\epsilon_{k'}),
  \label{DOS-def}
\end{eqnarray}
$\delta(\epsilon)$ is the Dirac delta function and $L$ is the
linear size of the bulk. For the dispersion (\ref{disp}), the DOS
is
\begin{subequations}
\begin{eqnarray}
  &&
  \nu(k) =
  \nu_1(\epsilon_k)+
  \nu_2(\epsilon_k),
  \label{DOS=DOS1+DOS2-res}
  \\
  &&
  \nu_1(\epsilon) =
  \vartheta(|\epsilon|-\epsilon_q)~
  \vartheta(\epsilon_0-|\epsilon|)~
  \frac{\rho_c |\epsilon|
        g_1(\epsilon)}
       {2\sqrt{\epsilon^2-\epsilon_q^2}},
  \label{DOS1-res}
  \\
  &&
  \nu_2(\epsilon) =
  \vartheta(|\epsilon|-\epsilon_q)~
  \frac{\rho_c |\epsilon|
        g_2(\epsilon)}
       {2\sqrt{\epsilon^2-\epsilon_q^2}},
  \label{DOS2-res}
\end{eqnarray}
where
\begin{eqnarray}
  &&
  \rho_c =
  \frac{v}{4\pi^2B^2\hbar^3}~
  \sqrt{Bm-\frac{1}{2}},
  \label{rho-c-def}
  \\
  &&
  g_i(\epsilon) =
  \sqrt{1+
        \big(-1\big)^i~
        \sqrt{\frac{\epsilon^2-\epsilon_q^2}
       {\epsilon_0^2-\epsilon_q^2}}},
  \label{gi-def}
\end{eqnarray}
  \label{subeqs-DOS-res}
\end{subequations}
$i=1,2$. Here $g_1(\epsilon)$ and $g_2(\epsilon)$ correspond to
the two solutions of the equation $\epsilon_k=\epsilon$.

For pure s wave scattering, $\eta_0(k)=\delta_S(|\epsilon_k-\epsilon_F|)$
and $\eta_l=0$ for $l\neq0$. Then Eq. (\ref{relax-time-phases}) can be
written as,
\begin{eqnarray}
  \frac{1}{\tau_{\mathrm{tr}}(k)} =
  \frac{2c_{\mathrm{imp}}}
       {\pi\hbar\nu(k)}~
  \sin^2\delta_S(|\epsilon_k-\epsilon_F|).
  \label{relax-time-phase-delta}
\end{eqnarray}
The scattering phase $\delta_S(\omega)$ is given by
Eq. (\ref{delta-BA}).
For $\omega=|\epsilon-\epsilon_F|\ll{T_K}$,
$\delta_S(\omega)$ is given by eq. (\ref{delta-S=1/2}) for
$S=\frac{1}{2}$ and by eq. (\ref{delta-S=1-asymptot}) for
$S=1$.

\noindent
{\textbf{Zero temperature resistivity}}:
As $T \to 0$ we can write
\begin{eqnarray}
  -\frac{\partial f(\epsilon)}
        {\partial\epsilon}
  &=&
  \delta(\epsilon-\epsilon_F).
  \label{df-T=0}
\end{eqnarray}
Substituting Eq. (\ref{df-T=0}) into Eq. (\ref{conduct-def}) and
taking into account Eqs. (\ref{group-velocity}),
(\ref{delta-BA}) and (\ref{subeqs-DOS-res}), we get
\begin{eqnarray*}
  \sigma^{(0)}(0) &=&
  \frac{\pi e^2}{3n_{\mathrm{imp}}}
  \int\frac{d^3\mbfk}{\big(2\pi\big)^3}~
  \delta(\epsilon_k-\epsilon_F)~
  v_k^2~
  \nu(k).
\end{eqnarray*}
When the Fermi energy is constrained according to
$$
  \veps_q<\epsilon_F<\veps_0,
$$
the equation $\veps_k=\epsilon_F$ has two solutions, $k=k_1$
and $k=k_2$, (see Fig. \ref{Fig-disp}). The equation
for the zero-temperature resistivity then takes the form,
\begin{eqnarray}
  &&
  \rho^{(0)}(0) =
  \frac{1}{\sigma^{(0)}(0)},
  \ \ \
  \sigma^{(0)}(0) =
  \sigma_{1}^{(0)}(0)+
  \sigma_{2}^{(0)}(0),
  \nonumber \\ &&
  \sigma_{i}^{(0)}(0) =
  \frac{\pi e^2
        \nu_i^2(\epsilon_k)
        v_{k_i}^2}
       {3c_{\mathrm{imp}}},
  \ \ \
  i=1,2.
  \label{conductivity-T=0-res}
\end{eqnarray}

\noindent
{\textbf{Finite temperature resistivity}}: When the temperature is
below the Kondo temperature $T_K$, the conductivity
(\ref{conduct-def}) can be written as,
\begin{eqnarray*}
  \sigma^{(0)}(T) &=&
  \frac{e^2}{3\pi^2}
  \int\limits_{0}^{q}
  \frac{k^2dk}{4T\cosh^2(\frac{\epsilon_k-\epsilon_F}{2T})}~
  v_k^2
  \tau_{\mathrm{tr}}(k)+
  \nonumber \\ &+&
  \frac{e^2}{3\pi^2}
  \int\limits_{q}^{\infty}
  \frac{k^2dk}{4T\cosh^2(\frac{\epsilon_k-\epsilon_F}{2T})}~
  v_k^2
  \tau_{\mathrm{tr}}(k).
\end{eqnarray*}
Substitute into each of integrals a factor unity,\\
$  \int{d\epsilon}~
  \delta(\epsilon-\veps_k)$=$1,$
and taking into account that for $k<q$ or $k>q$, the equation
$\veps_k=\epsilon$ has a single solution, $k_1(\epsilon)$ or
$k_2(\epsilon)$ (see Fig. \ref{Fig-disp} for illustration), we can
write
\begin{eqnarray}
  \sigma^{(0)}(T) &=&
  \frac{2e^2}{3}
  \int\limits_{\epsilon_q}^{\epsilon_0}
  \frac{\nu_1(\epsilon)
        v_1^2(\epsilon)
        \tau_{\mathrm{tr}}(k_1(\epsilon))
        d\epsilon}
       {4T\cosh^2(\frac{\epsilon-\epsilon_F}{2T})}+
  \nonumber \\ &+&
  \frac{2e^2}{3}
  \int\limits_{\epsilon_q}^{\infty}
  \frac{\nu_2(\epsilon)
        v_2^2(\epsilon)
        \tau_{\mathrm{tr}}(k_2(\epsilon))
        d\epsilon}
       {4T\cosh^2(\frac{\epsilon_k-\epsilon_F}{2T})},
  \label{conduct-relax}
\end{eqnarray}
where
$$
  v_i(\epsilon) \equiv
  v_{k_i(\epsilon)},
  \ \ \ \ \
  k_i(\epsilon)=
  q g_i(\epsilon),
  \ \ \ \ \
  i=1,2,
$$
and the functions $g_i(\epsilon)$ are given by Eq. (\ref{gi-def}).
\noindent
Taking into account Eqs. (\ref{relax-time-phase-delta}) and
(\ref{DOS=DOS1+DOS2-res}), we can write the conductivity
(\ref{conduct-relax}) as,
\begin{eqnarray}
  \sigma^{(0)}(T) &=&
  \frac{\pi \hbar e^2}{3 c_{\mathrm{imp}}}
  \int\limits_{\epsilon_q}^{\epsilon_0}
  \frac{\nu_1^2(\epsilon)
        v_1^2(\epsilon)
        \sin^{-2}\big(\delta_{S}(|\epsilon-\epsilon_F|)\big)
        d\epsilon}
       {4T\cosh^2(\frac{\epsilon-\epsilon_F}{2T})}+
  \nonumber \\ &+&
  \frac{\pi \hbar e^2}{3 c_{\mathrm{imp}}}
  \int\limits_{\epsilon_q}^{\infty}
  \frac{\nu_2^2(\epsilon)
        v_2^2(\epsilon)
        \sin^{-2}\big(\delta_{S}(|\epsilon-\epsilon_F|)\big)
        d\epsilon}
       {4T\cosh^2(\frac{\epsilon-\epsilon_F}{2T})}.
  \nonumber \\
  \label{conduct-phases-delta}
\end{eqnarray}
The integrands on the right hand side of Eq.
(\ref{conduct-phases-delta}) have a factor
$\cosh^{-2}(\frac{\epsilon-\epsilon_F}{2T})$ which is equal to 1
for $\epsilon=\epsilon_F$ and rapidly vanishes for
${\epsilon-\epsilon_F}\gg{T}$. The behaviour of the other factors,
$\nu_i^2(\epsilon)$ and $v_i^2(\epsilon)$ ($i=1,2$), depends on
the ratio $(\epsilon_F-\veps_q)/T$. When
$T\ll\epsilon_F-\veps_q$, then $\nu_i^2(\epsilon)$ and
$v_i^2(\epsilon)$ change slowly within the interval
$|\epsilon-\epsilon_F|\lesssim{T}$ and can be safely replaced by
$\nu_i^2(\epsilon_F)$ and $v_i^2(\epsilon_F)$. In what follows,
we will assume the inequality
$T_K\ll\epsilon_F-\veps_q$.\cite{TI3D-paper1} Then for $T<T_K$
the conductivity (\ref{conduct-phases-delta}) takes the form,
\begin{eqnarray}
  \sigma^{(0)}(T) &=&
  \sigma^{(0)}(0)
  \int\limits_{-\infty}^{\infty}
  \frac{\sin^{-2}\big[\delta_{S}(|\epsilon|)\big]
        d\epsilon}
       {4T\cosh^2(\frac{\epsilon}{2T})}.
  \label{conductivity-Tneq0-res}
\end{eqnarray}
(The limits of the integration can safely be changed to $\mp \infty$).
Finally, the impurity resistivity is,
\begin{eqnarray}
  \rho^{(0)}(T) =
  \rho^{(0)}(0)
  \left\{
       \int\limits_{-\infty}^{\infty}
       \frac{\sin^{-2}\big(\delta_{S}(|\epsilon|)\big)
             d\epsilon}
            {4T\cosh^2(\frac{\epsilon}{2T})}
  \right\}^{-1},
  \label{resist-Tneq0-res}
\end{eqnarray}
where $\rho^{(0)}(0)=1/\sigma^{(0)}(0)$ is the zero temperature
impurity resistivity, see Eq.~(\ref{conductivity-T=0-res}).

\noindent
{\textbf{Contribution of $\tilde{j}_{f}$ to the resistivity}}: The impurity
resistivity for the KE with the SO(3) symmetry
is given by Eq. (\ref{resist-Tneq0-res}). That for
the KE with SO(4) symmetry is contributed also
by the interaction of the $f$-impurity with the local Fermi
liquid. In order to derive this contribution, we apply
the perturbation poor man's scaling technique,
\begin{eqnarray}
  \delta\rho(T) &=&
  \frac{3R_0}{4}~
  \tilde{j}_{f}^{2}(T),
  \label{resist-Tneq0-correction}
\end{eqnarray}
where $\tilde{j}_{f}(T)$ is given by Eq. (\ref{tilde-jf-solution}).
The constant $R_0$ is,
\begin{eqnarray}
  R_0 &=&
  \frac{3 \pi c_{\mathrm{imp}}}
       {\hbar e^2 \rho_{\mathrm{c}}^{2}}~
  \frac{1}{v_{1}^{2}+v_{2}^{2}},
  \label{R0-def}
\end{eqnarray}
where
$$
  v_i ~=~
  v_{k_i},
  \ \ \ \ \
  i=1,2,
$$
$k_i$ are two solutions of the equation $\veps_k=\epsilon_F$,
see Fig. \ref{Fig-disp}.
Hence, the resistivity for the SO(4) KE reads,
\begin{eqnarray}
  \rho(T) &=&
  \rho^{(0)}(T)+
  \delta\rho(T),
  \label{resist-Tneq0-SO(4)}
\end{eqnarray}
where $\rho^{(0)}(T)$ is given by Eq. (\ref{resist-Tneq0-res}).
\begin{figure}[htb]
\centering
\includegraphics[width=60 mm,angle=0]{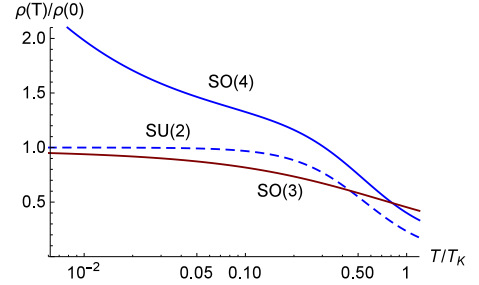}
\caption{\footnotesize
  {\color{blue}(Color online)}
  Resistivity (\ref{resist-Tneq0-res}) as a function of temperature
  for the SO(3) and SU(2) symmetries (solid red and dashed blue
  curves), and the resistivity (\ref{resist-Tneq0-SO(4)}) for
  the SO(4) KE (solid blue curve).}
 \label{Fig-resist-CFT}
\end{figure}
The resistivity (\ref{resist-Tneq0-res}) as a function of temperature
is shown in Figure \ref{Fig-resist-CFT} for the SO(4), SO(3) and SU(2)
symmetries. Recall that as ${T}\ll{T_K}$, the
resistivity for the SU(2) symmetry is given by,
\begin{eqnarray}
  \rho^{(0)}(T) &\approx&
  \rho^{(0)}(0)~
  \Bigg\{
      1-
      \frac{\pi^2 T^2}{3 T_K^2}+
      O\bigg(\frac{T^4}{T_K^4}\bigg)
  \Bigg\}.
  \label{resist-SO4-low-T}
\end{eqnarray}
This behaviour is typical for the Fermi liquid zero temperature
fixed point. Deviation from Eq.~(\ref{resist-SO4-low-T}) in the curves for the SU(2) and SO(4)
KE is due to the Kondo interaction between the $f$-impurity
and the local Fermi liquid.
The temperature dependence of the resistivity for the
SO(3) symmetry is rather distinct. Indeed, for ${T}\ll{T_K}$, we
can write
\begin{eqnarray}
  \rho^{(0)}(T) &\approx&
  \rho^{(0)}(0)~
  \Bigg\{
      1-
      \frac{\pi^2}{16 \ln^2\big(\frac{T_K}{2T}\big)}+
  \nonumber \\ && +
      O
      \Bigg(
           \ln^{-3}
           \bigg(
                \frac{T_K}{T}
           \bigg)
      \Bigg)
  \Bigg\}.
  \label{resist-SO3-low-T}
\end{eqnarray}
The singular temperature dependence of the resistance is the
result of the singular energy dependence of the scattering phase
(\ref{delta-BA}).

\section{Conclusions}
  \label{sec-conclusions}
The present work is motivated by the quest to elucidate
the elusive SFL behavior within bulk materials. This
property, associated with USKE, has so far been
observed only in electron tunneling experiments 
through quantum dots\cite{USKE-exp-PRL2009,%
USKE-exp-Sci2010,USKE-exp-PRL2015}, where
it is demonstrated that the tunneling conductance
$G$ approaches the unitary limit as $T \to 0$, but
the derivative $dG/dT$ diverges logarithmically. 
Elucidating SFL in metals based on the USKE
(e.g iron immersed in cupper) is not obvious
because magnetic impurities with high spin
give rise to Coqblin-Schrieffer type of scattering
which falls on a regular Fermi liquid fixed point
\cite{Hewson-book}. Here we achieved this goal
by analyzing Kondo scattering of electrons off
an Anderson impurity in a 3D topological insulator
with an ``inverted Mexican hat" band dispersion. 
We used the fact that the interplay between
the Anderson impurity and its induced in-gap
bound state results in self-screened Kondo effect
or in the Kondo effect with SO(4) or SO(3) dynamical
symmetries. Using the conformal field theory technique,
we have calculated the low temperature ($T \ll T_K$)
dependence of the specific heat, the magnetic
susceptibility and the electric resistivity of the impurity
for both screened and under-screened Kondo effect.
Physical properties of the impurity for the under-screened
case demonstrate zero temperature singularity which
corresponds to a singular Fermi liquid phase. In addition
to the SFL behavior exposed here, this system exposes
an interesting screening mechanism where the impurity
$f$ is screened by the quasi-particles of the Fermi sea 
that is formed when the band electrons and the $d$
impurity form a singlet state at $T<T_K$.

\noindent{\bf Acknowledgements}: We acknowledge support by HKRGC
through grant HKUST03/CRF09. The research of I.K., T.K. and Y.A. is
partially supported by grant 400/12 of the Israel Science
Foundation (ISF).

\appendix

\section{Anderson Model and Scaling Equations}
 \label{appendix}
Part of the material presented in this appendix
has already been developed in our previous
paper \cite{TI3D-paper1} where we analyzed
the same system in the weak coupling regime. 
It is included here for the sake of self-consistency.
In the first part we write down the bare Anderson
model and specify the peculiarities resulting from
the special form of the band structure. In the second
part we elaborate on the scaling equations 
and RG flow for the SO(4) dynamical symmetry.

\subsection{Anderson Model}
 \label{append-model}
\begin{figure}[htb]
\centering \subfigure[]
  {\includegraphics[width=50mm,angle=0]
  {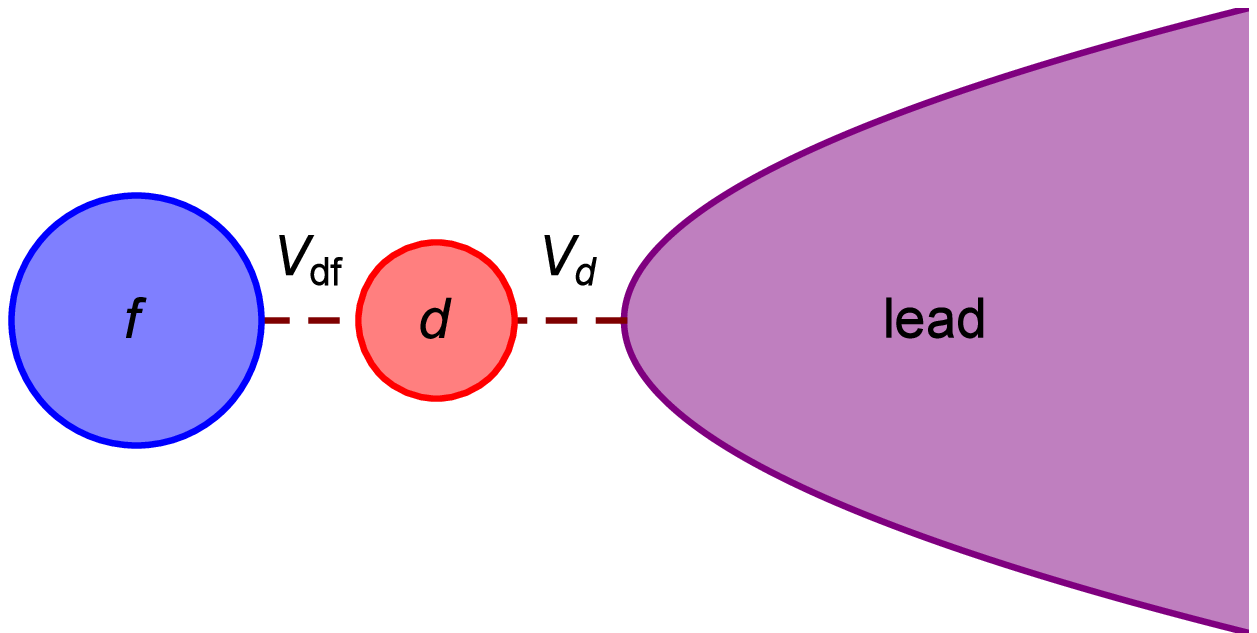}}
\subfigure[]
  {\includegraphics[width=50mm,angle=0]
  {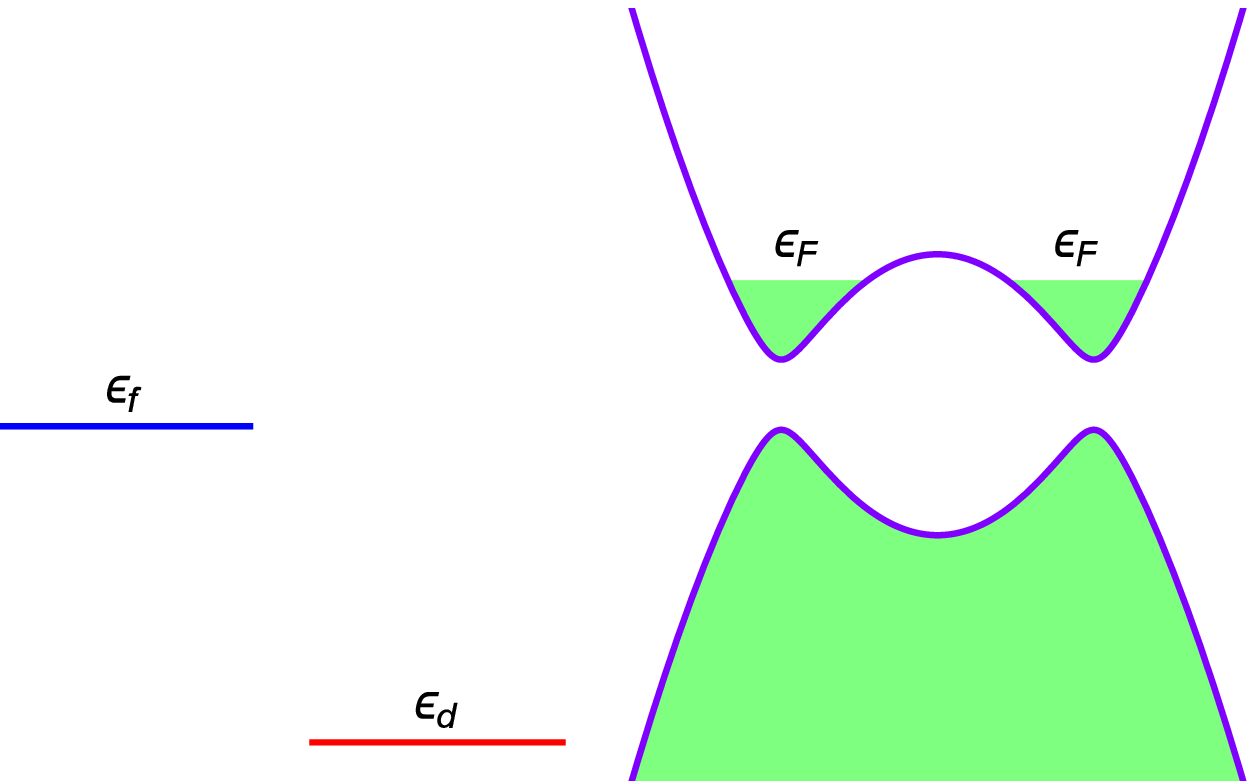}}
 \caption{\footnotesize
   {\color{blue}(Color online)}
   {\textbf{Panel (a)}}: Illustration of the Anderson model consisting
   of a band, $d$- and $f$-impurities.
   The hybridization rates of the $d$-impurity with
   the band and the $f$-impurity are $V_d$ and $V_{df}$,
   respectively.
   {\textbf{Panel (b)}} Energy dispersion (\ref{disp}),
   and energy levels $\epsilon_d$ and $\epsilon_f$.
   Here $\epsilon_F$ denotes the Fermi energy. All
   the energy levels of the band below the Fermi energy
   are occupied (green area), whereas the white area
   above $\epsilon_F$ denotes unoccupied levels.}
 \label{Fig-lead-d-f}
\end{figure}

The system considered here is schematically displayed in Fig. \ref{Fig-lead-d-f}a. It
consists of a topological insulator whose energy band has the
``inverted mexican hat'' structure, and an Anderson impurity
immersed in its bulk
(denoted as $d$-impurity). Potential scattering of
electrons on the impurity result in the formation a mid-gap localized
energy level \cite{TI3D-paper1}, denoted hereafter as an
$f$-impurity.
The effective Hamiltonian of the system is,
\begin{eqnarray}
  H &=&
  H_0+
  H_c+
  H_t.
  \label{H-tot-def}
\end{eqnarray}
Here the first term, $H_0$, is the Hamiltonian (\ref{H-ti-def})
describing electrons in the bulk of the TI.

The second term on the right hand side of eq. (\ref{H-tot-def}) is
the Hamiltonian of the isolated CQI, composed of the
$d$- and $f$-levels,
\begin{subequations}
\begin{eqnarray}
  H_c &=&
  H_d+H_f+H_{df}.
  \label{HC-def}
\end{eqnarray}
Here $H_d$ and $H_f$ are the (atomic) Hamiltonians of the $d$- and
$f$-impurities, and $H_{df}$ describes the hybridization between them,
\begin{eqnarray}
  H_d &=&
  \epsilon_d
  \sum_{\sigma}
  n_{d\sigma}+
  U_dn_{d\up}n_{d\down},
  \label{Hd}
  \\
  H_f &=&
  \epsilon_f
  \sum_{\sigma}
  n_{f\sigma}+
  U_f
  n_{f\up}
  n_{f\down},
  \label{Hf-def}
  \\
  H_{df} &=&
  V_{df}
  \sum_{\sigma}
  \left(
       f^{\dag}_{\sigma}
       d_{\sigma}+
       d^{\dag}_{\sigma}
       f_{\sigma}
  \right),
  \label{Hdf-def}
\end{eqnarray}
where $\epsilon_d$ or $\epsilon_f$ is the $d$- or $f$-impurity
energy level and $U_d$ or $U_f$ is the interaction between
electrons on the impurity.
Here $n_{d\sigma}=d_{\sigma}^{\dag}d_{\sigma}$,
$n_{f\sigma}=f_{\sigma}^{\dag}f_{\sigma}$, $d_{\sigma}^{\dag}$ or
$d_{\sigma}$ is the creation or annihilation operator of electron
on the $d$-level, $f_{\sigma}^{\dag}$ or $f_{\sigma}$ is the
creation or annihilation operator of electron on the $f$-level.
  \label{subeqs-H-complex-impurity-def}
\end{subequations}

The last term on the right hand side of eq. (\ref{H-tot-def}),
$H_t$, is the hybridization between the Anderson impurity $d$ and
the band electrons,
\begin{equation}
  H_t^{(0)} =
  V_d
  \sum_{{\mathbf{k}},\nu,\sigma}
  \left(
       \gamma_{\nu {\bf{k}}\sigma}^{\dag}
       d_{\sigma}+
       d^{\dag}_{\sigma}
       \gamma_{\nu {\bf{k}}\sigma}
  \right).
 \label{Ht}
\end{equation}
Note that hybridization prevails only between the band electrons and the $d$ impurity.
The $f$-level is formed as a result of the potential scatering
of the conduction band electron on the $d$-impurity
\cite{TI3D-paper1}. Therefore, hybridization of $f$-impurity
and the band electrons is absent.

\noindent
{\bf Energy scales}:
Few words about energy scales are in order: Unless otherwise
specified, we shall assume that $U_d\rightarrow\infty$ and
\begin{equation}
  \label{escale}
  \epsilon_F-D_0 <
  \epsilon_d \ll
  \epsilon_f <
  \epsilon_F <
  \epsilon_f+U_f\ll\epsilon_F+D_0,
\end{equation}
where $D_0$ (the initial bandwidth) is the highest energy cutoff,
and $\epsilon_F$ is the Fermi energy (see Figure
\ref{Fig-lead-d-f}).  In the calculations below we use $\epsilon_F=2\veps_q$,
$\epsilon_f\approx-\veps_q$, $U_f=5\veps_q$ and
$\epsilon_d=-80\veps_q$.

\noindent
{\bf The eigenstates of $H_c $},
equation  (\ref{HC-def}), are specified by the configuration
numbers $(N_d,N_f)$ indicating the number of electrons on the
levels $d$ and $f$. With energy scales specified in
Eq.~(\ref{escale}), the ground state (GS) has $N_d=N_f=1$ and there are four
possible states, a spin-singlet state $|S\rangle$ and three
spin-triplet states $|T_m\rangle$ ($m=0,\pm1)$. The singlet energy
is modified when $V_{df}\ne0$ while the triplet energy is
unaffected. Explicitly,
\begin{subequations}
\begin{eqnarray}
  &&
  |S\rangle =
  \left\{
       \frac{\alpha_S}{\sqrt{2}}
       \Big(
           d_{\up}^{\dag}
           f_{\down}^{\dag}-
           d_{\down}^{\dag}
           f_{\up}^{\dag}
       \Big)-
       \beta_f
       f_{\up}^{\dag}
       f_{\down}^{\dag}
  \right\}
  |0\rangle,
  \label{singlet-def}
  \\
  \nonumber
  \\
  &&
  \begin{array}{l}
  \displaystyle
  |T_1\rangle=
  d_{\up}^{\dag}
  f_{\up}^{\dag}
  |0\rangle,
  \ \ \ \ \
  |T_{-1}\rangle=
  d_{\down}^{\dag}
  f_{\down}^{\dag}
  |0\rangle,
  \\
  \\
  \displaystyle
  |T_{0}\rangle =
  \frac{1}{\sqrt{2}}
  \Big\{
      d_{\up}^{\dag}
      f_{\down}^{\dag}+
      d_{\down}^{\dag}
      f_{\up}^{\dag}
  \Big\}
  |0\rangle,
  \end{array}
  \label{triplet-def}
\end{eqnarray}
  \label{subeqs-singlet-triplet}
\end{subequations}
where
\begin{eqnarray*}
  &&
  \varepsilon_{S} =
  \epsilon_d+
  \epsilon_f-
  \frac{2V_{df}^2}{\Delta_f},
  \ \ \ \ \
  \varepsilon_{T} =
  \epsilon_d +
  \epsilon_f,
  \\
  &&
  \alpha_S =
  \sqrt{1-\beta_f^2},
  \ \ \ \ \
  \beta_f =
  \frac{\sqrt{2}V_{df}}{\Delta_f},
  \\
  &&
  \Delta_f =
  \epsilon_f-
  \epsilon_d+
  U_f.
\end{eqnarray*}
In the absence of hybridization of d-electron with the band
electrons (i.e., when $V_d=0$), $\veps_S<\veps_T$: As expected, the
singlet state has lower energy than the triplet state.

\subsection{Scaling Equations for the SO(4) Dynamical Symmetry}
  \label{append-scaling-SO(4)}

In order to perform the RG analysis considered in
Sec. \ref{sec-second-TK} [see Eq. (\ref{eq-tilde-jf})], we need
the effective coupling $j_f(T_{K_4})$ at the effective bandwidth
equal to $T_{K_4}$. To derive $j_f(T_{K_4})$, we apply the weak
coupling RG analysis for the SO(4) KE. In the the weak coupling
regime, the  scaling equations for the dimensionless couplings
$j_d=J_d\rho_0$ and $j_f=J_f\rho_0$ must proceed to third order
in these parameters \cite{Hewson-book}, that is,
\begin{subequations}
\begin{eqnarray}
  \frac{\partial j_d}{\partial \ln D}
  &=&
  -j_d^2+
  j_d
  \big(
      j_d^2+
      j_f^2
  \big),
  \label{eq-jd}
  \\
  \frac{\partial j_f}{\partial \ln D}
  &=&
  -j_f^2+
  j_f
  \big(
      j_d^2+
      j_f^2
  \big).
  \label{eq-jf}
\end{eqnarray}
  \label{subeqs-scaling-eqs}
\end{subequations}
The initial values of $j_d$ and $j_f$ at $D-D_{ii}$ are
$J_d\rho_{\mathrm{c}}$ and $J_f\rho_{\mathrm{c}}$,
where $J_d$ and $J_f$ are given by Eq. (\ref{JK-def}).
\begin{figure}[htb]
\centering
\includegraphics[width=60 mm,angle=0]
   {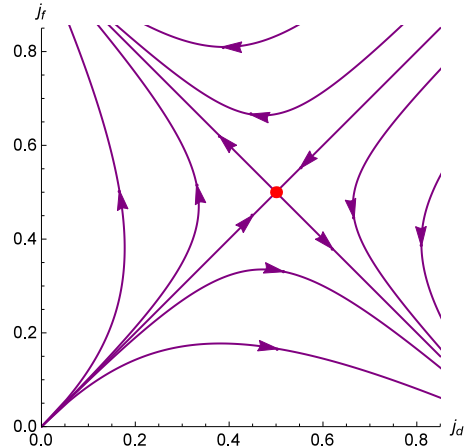}
 \caption{\footnotesize
   {\color{blue}(Color online)}
   Scaling of $j_d$ and $j_f$.}
 \label{Fig-scaling}
\end{figure}
Scaling of $j_d$ and $j_f$ is shown in the flow diagram,
Fig. \ref{Fig-scaling}. The red dot denotes the two-channel fixed point
$j_{d}^{*}=j_{f}^{*}=1/2$. We are interested in the solution
of these equations under the following inequalities:
$$
  {j}_{f}(D_{ii}) ~\ll~
  {j}_{d}(D_{ii}) ~\ll~ 1,
$$
so that we are far away from the two-channel fixed point. In this
case $j_d(D)$ increases when $D$ decreases, whereas $j_f(D)$
decreases and goes to zero when $D$ vanishes. The solution of
the scaling equation (\ref{eq-jd}) can be approximated as,
\begin{eqnarray}
  j_d(D) &=&
  \frac{1}
       {\displaystyle
        \ln
        \bigg(
             \frac{D}{T_{K_4}}
        \bigg)},
  \label{jd-solution-scaling-limit}
\end{eqnarray}
where the scaling invariant, the Kondo temperature $T_{K_4}$, is
\begin{eqnarray}
  T_{K_4} =
  D_{ii}~
  \exp
  \bigg\{
       -\frac{1}{J_d\rho_{\mathrm{c}}}
  \bigg\}.
  \label{TK4-def}
\end{eqnarray}
The solution (\ref{jd-solution-scaling-limit}) is valid just for
${D}\gg{T}_{K_4}$. A more satisfactory approximation
is given in Ref. \cite{Hewson-book},
\begin{eqnarray}
  j_d(D) =
  \sqrt{\frac{8}{3\pi^2}}~
  \sqrt{
       1-
       \frac{\displaystyle
             \ln
             \bigg(
                  \frac{D}{T_{K_4}}
             \bigg)}
            {\displaystyle
             \sqrt{\ln^2\bigg(\frac{D}{T_{K_4}}\bigg)+
                   \frac{3\pi^2}{4}}}}.
  \label{jd-solution-scaling-limit-beter}
\end{eqnarray}

The solution of the scaling equation (\ref{eq-jf}) is,
\begin{eqnarray}
  \frac{j_f(D)}{j_f(D_{ii})} &=&
  \exp
  \Bigg\{
       -\int\limits_{D}^{D_{ii}}
       \frac{d\epsilon}{\epsilon}~
       j_{d}^{2}(\epsilon)
  \Bigg\}.
  \label{jf-solution-integral-scaling-limit}
\end{eqnarray}
Taking into account eq. (\ref{jd-solution-scaling-limit-beter}),
we get
\begin{eqnarray*}
  \frac{j_f(D)}{j_f(D_{ii})} &=&
  \exp
  \Bigg\{
       \frac{8}{3\pi^2}
       \bigg(
            {\cal{F}}(D_{ii})-
            {\cal{F}}(D)
       \bigg)
  \Bigg\},
\end{eqnarray*}
where
\begin{eqnarray*}
  {\cal{F}}(D) &=&
  \sqrt{\ln^2\bigg(\frac{D}{T_{K_4}}\bigg)+
        \frac{3\pi^2}{4}}-
  \ln\bigg(\frac{D}{T_{K_4}}\bigg).
\end{eqnarray*}
When $D$ approaches $T_{K_4}$, $j_f(T_{K_4})$ can be
approximated as
\begin{eqnarray}
  j_f(T_{K_4}) &=&
  j_f\big(D_{ii}\big)~
  \exp
  \left\{
       \displaystyle
       -\frac{4}{\sqrt{3}\pi}
  \right\}
  ~\approx \nonumber \\ &\approx&
  0.5
  j_f\big(D_{ii}\big),
  \label{jf-solution-scaling-limit-1}
\end{eqnarray}
where we take into account that $D_{ii}\gg{T}_{K_4}$.


\end{document}